\documentclass[prl,twocolumn,amsmath,amssymb,showpacs]{revtex4} 
\usepackage{epsfig,amsmath}
\usepackage{amssymb,amsmath,wasysym}
\usepackage{bbm}	
\usepackage[normalem]{ulem} 

\newcommand{\beq}{\begin{equation}}
\newcommand{\eeq}{\end{equation}}
\newcommand{\beqarr}{\begin{eqnarray}}
\newcommand{\eeqarr}{\end{eqnarray}}

\newcommand{\KSum}{\mathop{\vcenter{\hbox{\huge$\mathrm{K}$}}}}
\newcommand{\cfracslash}[2]{\; \begin{array}{c}\multicolumn{1}{@{\,}c@{\,}|}{#1}\\ \hline \multicolumn{1}{|@{\,}c@{\,}}{#2} \end{array}}

\begin{document}

\title{Synchronization Transition in the Kuramoto Model with Colored Noise}
\author{Ralf T\"onjes}
\affiliation{Ochadai Academic Production, Ochanomizu University, Tokyo 112-8610, Japan}

\begin{abstract}
We present a linear stability analysis of the incoherent state in a system of globally coupled, identical phase oscillators subject to colored noise. In that we succeed to bridge the extreme time scales between the formerly studied and analytically solvable cases of white noise and quenched random frequencies.
\end{abstract}

\maketitle

The term Kuramoto Model refers to a class of nonlinear models which describe the dynamics of autonomous limit cycle oscillators by phase equations and interactions between them via coupling functions of phase differences. Since its original formulation \cite{Kuramoto75} it has been modified to include for instance other nonlinear effects, coupling topology or delayed coupling \cite{AceBon05}. An important property of these models is the existence of a transition to synchronization in large systems of coupled oscillators mediated through the opposing effects of attractive interaction and heterogeneity. Synchronization is a collective phenomenon where the phases of the oscillators become correlated leading to macroscopic oscillations or more complicated behavior \cite{Kuramoto84,MontBlaKu05,OttAnton08,PiRo08,ToKoMa10}. The Kuramoto Model is therefore able to reproduce a fundamental mechanism of self-organization in nature which is important for pattern formation, information processing and transport among others.
\\
In \cite{Kuramoto84} Kuramoto considers the case of all-to-all coupling where each oscillator couples equally strong to all other oscillators in the system. The Kuramoto phase equations for such a system are
\beq	\label{Eq:KM01}
	{\dot\vartheta_n} = \sigma\eta_n + \frac{1}{N} \sum_{m=1}^N g(\vartheta_m-\vartheta_n)
\eeq
where $\vartheta_n$ is the phase of the oscillator $n$, $\sigma\eta_n$ is an individual force which may be the natural frequency of the oscillator or a time dependent perturbation, $g(\Delta\vartheta)$ is a periodic coupling function of a phase difference and $N$ is the total number of oscillators. Disorder is realized through a distribution of random forces $\sigma\eta_n$, where $\sigma$ denotes the noise amplitude in units of coupling strength. When the forces are time independent the system models an ensemble of oscillators with nonidentical natural frequencies. For quenched random frequencies with unimodal distribution a continuous phase transition from an incoherent regime of evenly distributed phases to a regime of partial synchronization can be observed when $\sigma$ is changed. In fact, depending on the shape of the frequency distribution or the coupling function, even more complicated behavior is possible \cite{Kuramoto84,KoriKiss07,StrogatzOtt09}. If, on the other hand, the $\eta_n$ change very rapidly, they may be approximated by white noise. Again, a continuous phase transition is predicted as the noise strength is changed \cite{Kuramoto84}. These two analytically solvable cases mark the extremes of time scale separation between the dynamics of the oscillators and the fluctuations. In experiments, however, system parameters may drift at time scales comparable to the drift of the oscillator phase differences. Moreover, if the random forces $\sigma\eta_n$ are intrinsic to the system, for instance, in  phase coherent chaotic oscillators, or in a random network of identical phase oscillators \cite{Ott05,ToKoMa10}, the time scales are not necessarily separated. It is therefore of great interest to know how the critical coupling strength for the phase transition to partial synchronization for colored noise differs from the known values at quenched or white noise.
\\
Numerical investigations of that question have been carried out and qualitative answers have been given at selected parameters \cite{HuPeBag07}. However, an analytical solution to the problem has so far remained an open problem. Here we provide the solution in the two cases of the random telegraph process (TP) and the Ornstein-Uhlenbeck process (OU) as source of the colored noise. We find, that the type of the random process is essential for the transition point to synchronization, as can be expected from the rich behavior of the Kuramoto model with different quenched frequency distributions \cite{PiRo08,StrogatzOtt09,SakKura86}.
\section*{Evolution of Phase and Frequency Distribution}
%
%
In the thermodynamical limit $N\to\infty$ the system can be described by a density $p(\vartheta,\eta,t)$ of phases $\vartheta$ and forces $\eta$. The evolution of this density is given by a Fokker-Planck equation
\beq	\label{Eq:FPE01}
	\partial_t p ~=~ L_\vartheta\left[p\right] p + L_\eta p
\eeq
where we assume that the forces $\eta_n$ are independent, linear random processes described by a linear operator $L_\eta$, whereas the Fokker-Planck operator $L_\vartheta\left[p\right] \cdot = -\partial_\vartheta(\dot\vartheta\cdot)$ that acts on the phases depends on the mean field and is therefore a functional of the oscillator density $p$. Our strategy is to linearize Eq.~(\ref{Eq:FPE01}) around the stationary incoherent state $p(\vartheta,\eta,t)=p(\eta)/2\pi$ and look for a critical condition of the stability of the eigenmodes of $p$. If $L_\eta$ has a finite number of eigenmodes, as is the case for the random telegraph process, we will only have to solve a finite system of linear equations. This is not the case, however, when we consider the probably most applied case of the OU process. Then $L_\eta$ has an infinite but countable number of eigenmodes and we are faced to determine the stability of an infinite system of linear ODEs.
\\
Given the eigenvalues $\lambda_n$ and eigenmodes $\varphi_n(\eta)$ of $L_\eta$ with $L_\eta\varphi_n = \lambda_n\varphi_n$ and $0 = \lambda_0 > \textnormal{Re}\lambda_1 \dots$ we start by expanding $p(\vartheta,\eta)$ and $g(\Delta\vartheta)$ as
\beqarr	\label{Eq:ModesExpansion}
	p(\vartheta,\eta) 	&=& 	\frac{1}{2\pi}\sum_{k=-\infty}^{\infty} \sum_{n=0}^{\infty} z_{kn} e^{\textnormal{i}k\vartheta} \varphi_n(\eta),	\nonumber \\ \\
	g(\Delta\vartheta) 	&=&	\sum_{k=-\infty}^{\infty} g_{k}^* e^{-ik\Delta\vartheta}					\nonumber ~.
\eeqarr
The Kuramoto phase equation (\ref{Eq:KM01}) gives
\beqarr	\label{Eq:KM02}
	{\dot\vartheta} 	&=& 	\sigma\eta + \int_{-\pi}^{\pi} d\vartheta' \int_{-\infty}^{\infty} d\eta' g(\vartheta'-\vartheta) p(\vartheta',\eta') \nonumber \\ \\
				&=&	\sigma\eta + \sum_{k=-\infty}^{\infty} z_{k0}~ g_k^*~ e^{\textnormal{i}k\vartheta}	~.					\nonumber
\eeqarr
Inserting Eqs.~(\ref{Eq:ModesExpansion}) and (\ref{Eq:KM02}) into (\ref{Eq:FPE01}), we find for the modes $z_{kn}$ the nonlinear equation
\beq
	{\dot z}_{kn} ~=~ -\textnormal{i}k \sum_{l=-\infty}^{\infty} z_{l0} g_l^* z_{(k-l) n} ~-~ \textnormal{i}k\sigma \sum_{m=0}^{\infty} z_{km} M_{mn} + \lambda_n z_{kn}	
\eeq
with
\beq
	\eta\varphi_n(\eta) = \sum_{m=0}^{\infty} M_{nm} \varphi_m(\eta)		~.
\eeq
We remark that the celebrated ansatz of Antonsen and Ott \cite{OttAnton08} corresponding to $c_{k}(\eta) := \sum_n z_{kn}\varphi_n/\varphi_0 = a^k(\eta)$ ($a^{*|k|}$ for $k<0$) only leads to a closed nonlinear dynamic equation for $a(\eta)$ when all eigenvalues $\lambda_n$ are equal to zero, i.e., in the limit of quenched random frequencies.
\\
The order parameter $R=|\left\langle e^{i\vartheta} \right \rangle|$ is equal to the absolute value of $z_{10}$ which is zero when the phases are distributed uniformly in the incoherent state.
One can check that the incoherent distribution $p(\vartheta,\eta,t) = \varphi_0(\eta)/2\pi$ or $z_{kn}=\delta_{k0}\delta_{n0}$ is a stationary solution of the Fokker-Planck equation (\ref{Eq:FPE01}). Thus, keeping only the terms linear in the small quantities $z_{kn}$ we obtain linearized equations for the dynamics of the modes
\begin{figure}[!th] 
\center
 \includegraphics[width=4.1cm]{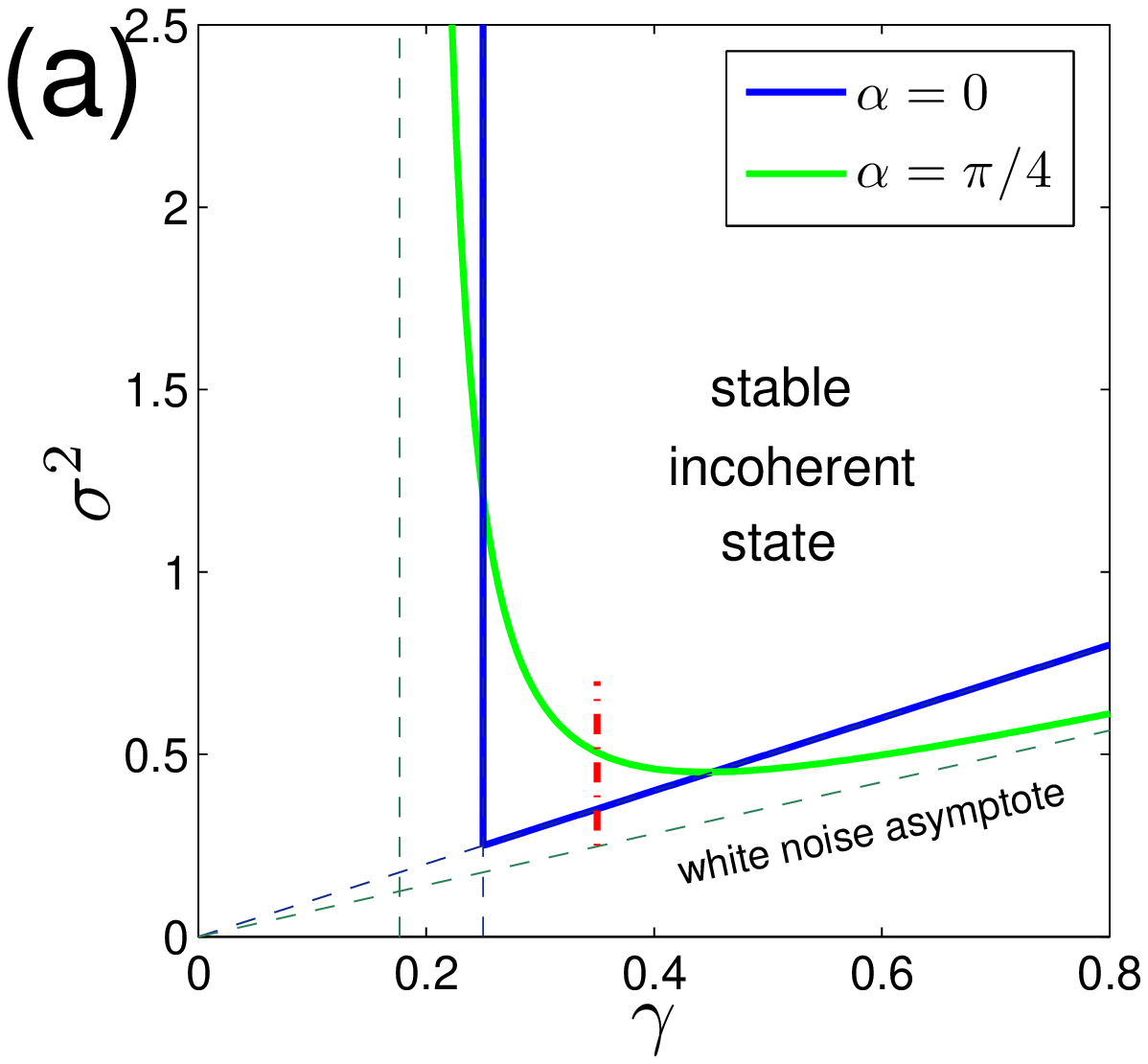}
 \includegraphics[width=4.1cm]{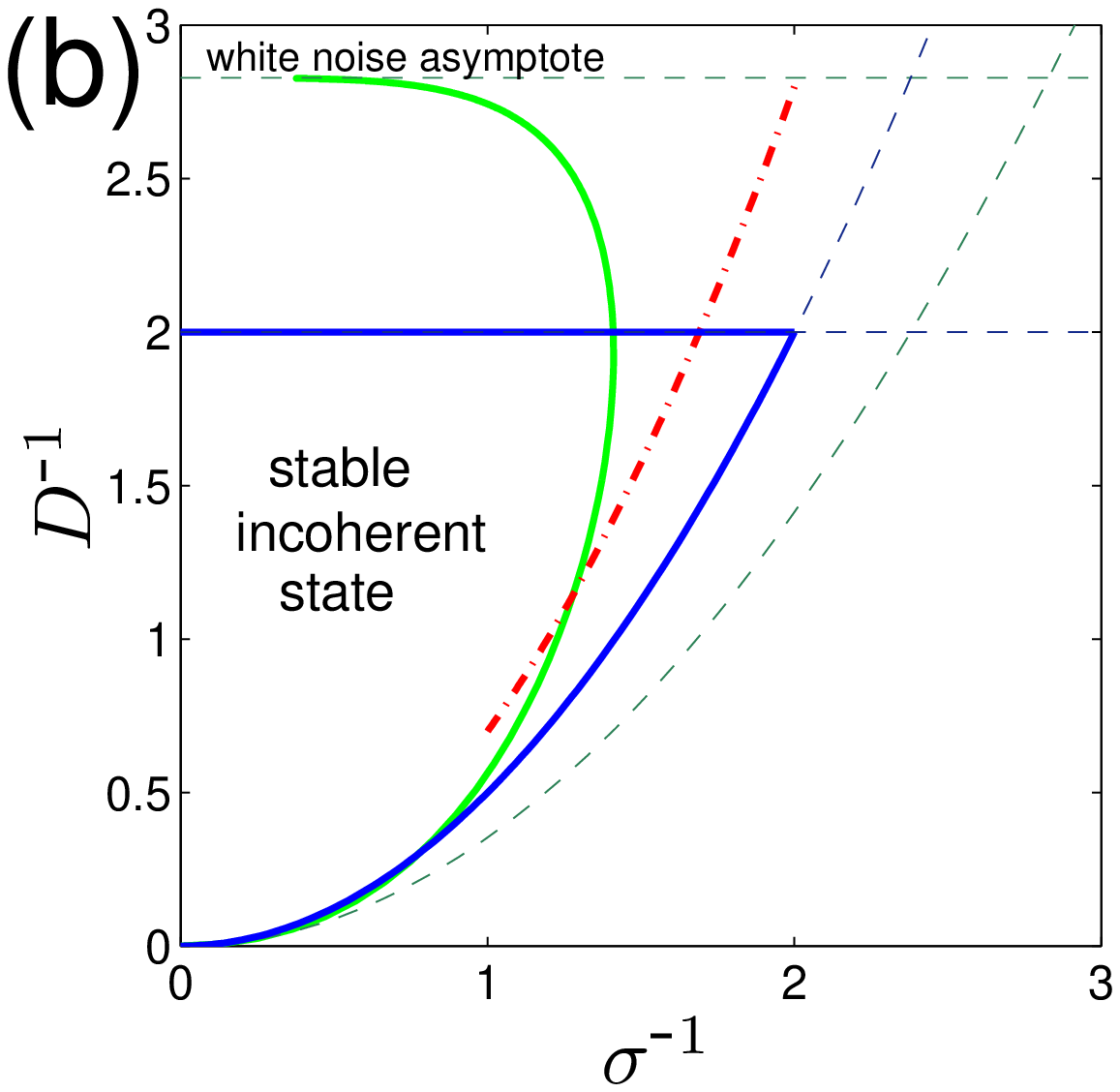}
 \includegraphics[width=4.1cm]{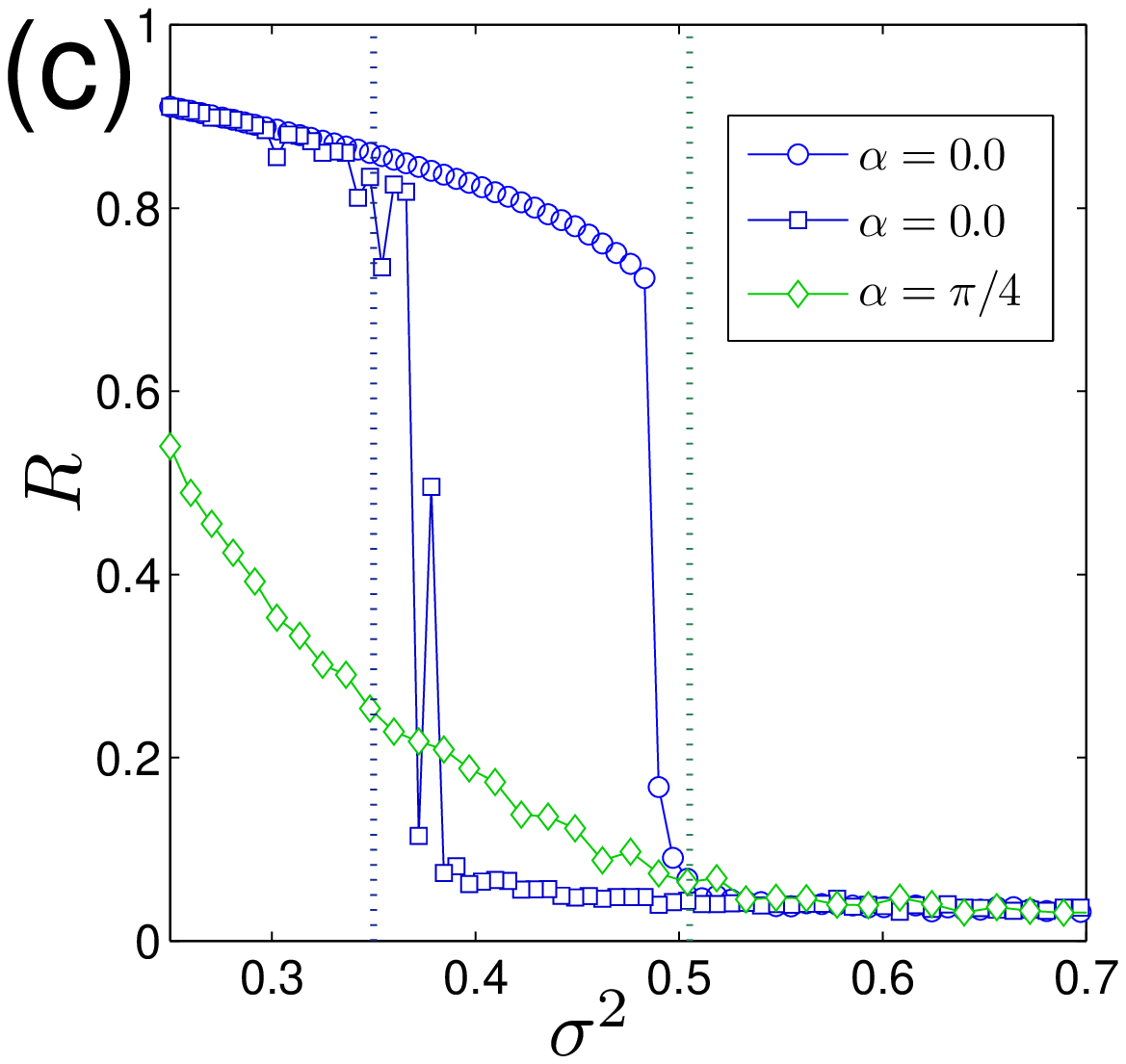}
 \includegraphics[width=4.1cm]{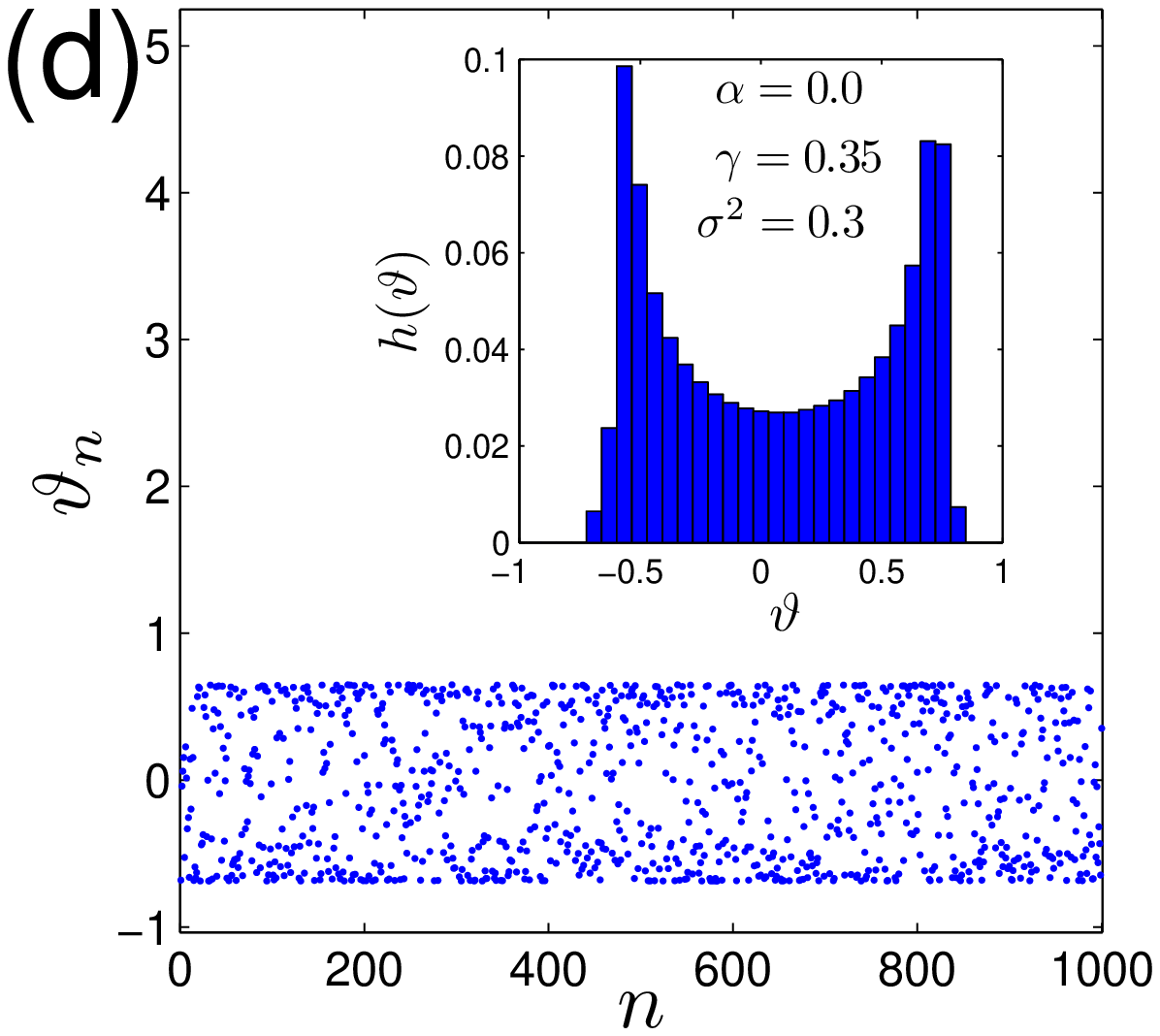}
\caption{\small (color online) Case of the random telegraph process. Subfigures (a) and (b) show the parameter regions of linear stability of the incoherent state for (a) switching rate $\gamma$ and amplitude $\sigma^2$ or (b) inverse amplitude $\sigma^{-1}$ and noise strength $D^{-1}$. Shown are the two cases $\alpha=0$ (bold dark/blue) and $\alpha=\pi/4$ (bold light/green).  The white noise limit is reached on the vertical axis of (b) where $\sigma^{-1}\to 0$ and $D^{-1}=\gamma/2\sigma^2=const$. Subfigure (b) can directly be compared to the case of Ornstein-Uhlenbeck type noise in Fig.~\ref{Fig:OUfig}.  The dashed lines are asymptotes obtained from Eq.~(\ref{Eq:TN_Hopf_Line}). Subfigure (c) shows the time-averaged  order parameter. The theoretical predictions are verified by direct numerical simulation in systems of size $N=5000$ at $\gamma=0.35$ using $\sigma^2$ as bifurcation parameter. The corresponding one dimensional curve is shown as dashed and dotted (red) line in (a) and (b). The vertical dotted lines mark the theoretical transition points. Hysteresis is observed for $\alpha=0.0$ (blue circles and squares). Subfigure (d) shows the phases of $1000$ oscillators in the partially synchronized state for $\alpha=0.0$, $\gamma=0.35$ and $\sigma^2=0.3$ with a phase histogram in the inset. In this state the oscillators are phase locked to the stationary mean field.}
\label{Fig:TNfig}
\end{figure}
\beqarr	\label{Eq:LinEqAll}
	{\dot z}_{k0}	&=&	- \textnormal{i}k \left( g_0^* + g_k^* \right) z_{k0} - \textnormal{i}k\sigma\sum_{m} z_{km} M_{m0},	\nonumber \\ \\
	{\dot z}_{kn}	&=&	\left( \lambda_n - \textnormal{i}k g_0^* \right) z_{kn} - \textnormal{i}k\sigma\sum_{m} z_{km} M_{mn}.	\nonumber
\eeqarr
The Fourier modes of the probability distribution, i.e. the $z_{kn}$ for different $k$, do not interact linearly with one another and can be studied separately. The term $-\textnormal{i}kg_0^*=-\textnormal{i}kg_0$ which appears as a self-interaction term for all eigenmodes can be neglected for the stability analysis since it is imaginary and only results in a bias to all frequencies. The incoherent state becomes unstable when the largest real part of the eigenvalues of the linear ODE (\ref{Eq:LinEqAll}) becomes positive. For a linear random process with a finite number of eigenmodes $\varphi_n(\eta)$ we only need to to determine the eigenvalues of a square matrix depending on the system parameters.
\section*{Random Telegraph Process}
%
%
Consider the Kuramoto model with sinusoidal coupling function $g(\Delta\vartheta)=\sin(\Delta\vartheta-\alpha)$, attracting ($\cos\alpha>0$) and with independent random forces $\eta_n\in\lbrace -1,1\rbrace$ which change sign as a dichotomous random Markov process with equal transition rate $\gamma$ between both values. Following the analysis in the previous section we find that the linear stability of the first Fourier mode at the incoherent state is determined by the eigenvalues of the matrix
\beq
	\mathit{J} = \left(
 	\begin{array}{cc}
 		\frac{1}{2}e^{\textnormal{i}\alpha} & \textnormal{i}\sigma \\
 		\textnormal{i}\sigma		      & -2\gamma
 	\end{array}
 	\right)		~.
\eeq
Given $\alpha$ and the flipping rate $\gamma$, necessary and sufficient conditions for stability are
\beq \label{Eq:TN_Hopf_Line}
	\sigma^2	>	\gamma\cos\alpha \left(1+\frac{\sin^2\alpha}{(4\gamma-\cos\alpha)^2}\right)	~,~	\cos\alpha	-	4\gamma 		< 0.
\eeq
The frequency $\Omega$ of the mean field at the bifurcation is $\Omega=2\gamma\sin\alpha/(4\gamma-\cos\alpha)$. Interestingly increasing $\alpha$ the incoherent state may actually become unstable in some regions of parameter space (see Fig.~\ref{Fig:TNfig}a,b) even though it is stable everywhere when $\alpha\to\pi/2$. To test our result we integrated Eq.~(\ref{Eq:KM01}) numerically with time steps determined by the random switching events and find it in good agreement with the prediction (Fig.~\ref{Fig:TNfig}c). Both the thermodynamic and the white noise limit are not accessible through this integration scheme since the time step is of order $O(1/N\gamma)$. However, the white noise limit with $D=2\sigma^2/\gamma$ is recovered from Eq.~(\ref{Eq:TN_Hopf_Line}) with $D_{\textnormal{cr}}=0.5 \cos\alpha$ and $\Omega=0.5 \sin\alpha$. From the linear stability analysis we are not able to predict whether the Hopf-Bifurcation is supercritical or subcritical. In fact, the simulations show either behavior in different parameter regions.
\section*{Ornstein-Uhlenbeck Process}
%
Instead of randomly switching between two values, we now consider i.i.d. random forces diffusing in the fashion of an OU process with Langevin equation
\beq
	{\dot\eta} = -\gamma\eta + \sqrt{2\gamma}\xi(t)
\eeq
and white noise $\left\langle\xi(t)\xi(t')\right\rangle = \delta(t-t')$. The rate $\gamma$ determines the time scale of the diffusion. To visualize both the white noise and the quenched noise limit it is of advantage to use the parameter $D=\sigma^2/\gamma$. Then the white noise limit with finite noise strength $D$ is reached as $\sigma^{-1}\to 0$ and quenched noise corresponds to $D^{-1}\to 0$.
The eigenvalues and eigenfunctions of the Fokker-Planck operater $L_\eta$ for an OU process are intimately related to those of the quantum harmonic oscillator \cite{Risken89}. One finds $\lambda_n=-\gamma n$ and $M_{mn} = \delta_{m n-1}\sqrt{n} ~+~ \delta_{m n+1}\sqrt{n+1}$. This turns the eigenvalue problem of Eq.~(\ref{Eq:LinEqAll}) into an infinite system of second order difference equations. 
\\
While determining the eigenvalues of the ODE (\ref{Eq:LinEqAll}) depending on $D$, $\sigma$ and $\alpha$ presents a major difficulty, this is actually not necessary. Instead we notice that at the transition to synchronization there is an imaginary eigenvalue $\textnormal{i}\Omega$, which gives us an implicit condition for the bifurcation line of the first Fourier mode ($k=1$)
\beqarr	\label{Eq:CritDiffEq}
	\textnormal{i}\Omega z_{10} &=& - \textnormal{i}  g_1^* z_{10} - \textnormal{i}\sigma z_{11}, 		\nonumber \\ \\
	\textnormal{i}\Omega z_{1n} &=& -\sigma^2 D^{-1} n z_{1n} -i \sigma\sqrt{n} z_{1 n-1} - \textnormal{i}\sigma\sqrt{n+1} z_{1n+1}. \nonumber
\eeqarr
At the transition $\Omega$ is the frequency of the emerging mean field. Denoting $\mu_n = -i  \sqrt{n+1} z_{1n+1}/z_{1n}$ these difference equations define a continued fraction 
\beqarr	\label{Eq:Kette01}
	&\mu_0 =  \textnormal{i}\left(g_1^*\sigma^{-1} + \Omega\sigma^{-1}\right) = \mu_0\left(D^{-1}\sigma,\Omega\sigma^{-1}\right),&	\nonumber \\ \\
	&\mu_0(x,y) = -\KSum_{n=1}^{\infty} \cfracslash{n}{nx+\textnormal{i}y}&		\nonumber
\eeqarr
where the dimensionless quantities $x=\gamma\sigma^{-1}=D^{-1}\sigma$ and $y=\Omega\sigma^{-1}$ relate the dynamical time scales in the system.
This equation for the critical condition is one of the main results of this paper. The complex function $\mu_0(x,y)$ can be calculated efficiently from Eq.~(\ref{Eq:Kette01}). Using a technique of Euler \cite{Euler1782}, one can find the continued fraction in terms of functions related to confluent hypergeometric functions of the first kind
\beqarr \label{Eq:SpecialRatio}
	\mu_0(x,y) &=& -\frac{1}{x}~\frac{_1f_1\left(2,x^{-1}\left(\textnormal{i}y+x^{-1}\right)+2,x^{-2}\right)}{_1f_1\left(1,x^{-1}\left(\textnormal{i}y+x^{-1}\right)+1,x^{-2}\right)},	\quad	\nonumber \\  \\
	_1f_1(a,b,z) &=& \int_{0}^1 du ~u^{a-1} (1-u)^{b-a-1} e^{zu} 	~.												\nonumber
\eeqarr
\begin{figure}[!tbh] 
\center
\includegraphics[width=8cm]{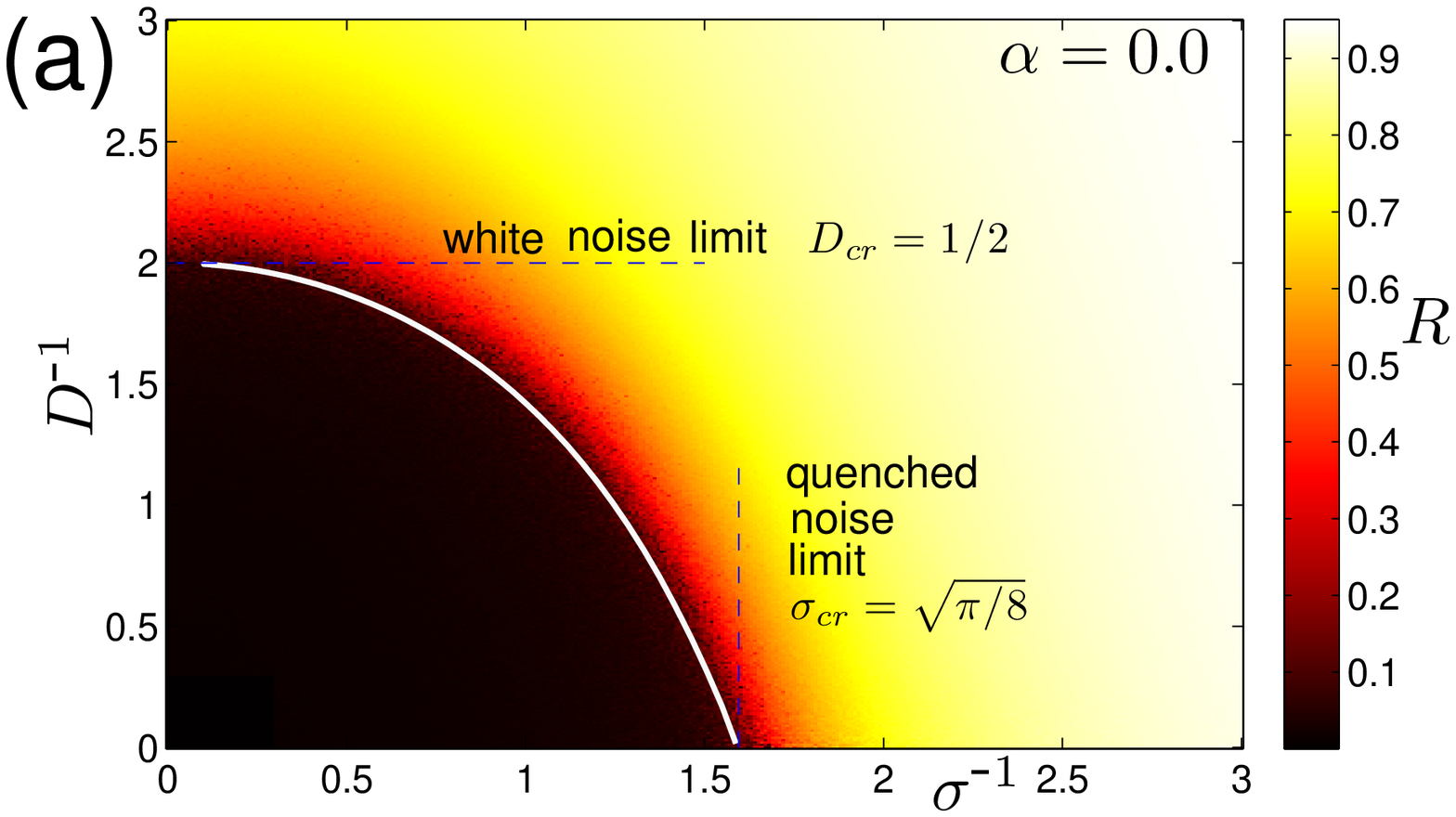}
\includegraphics[width=8cm]{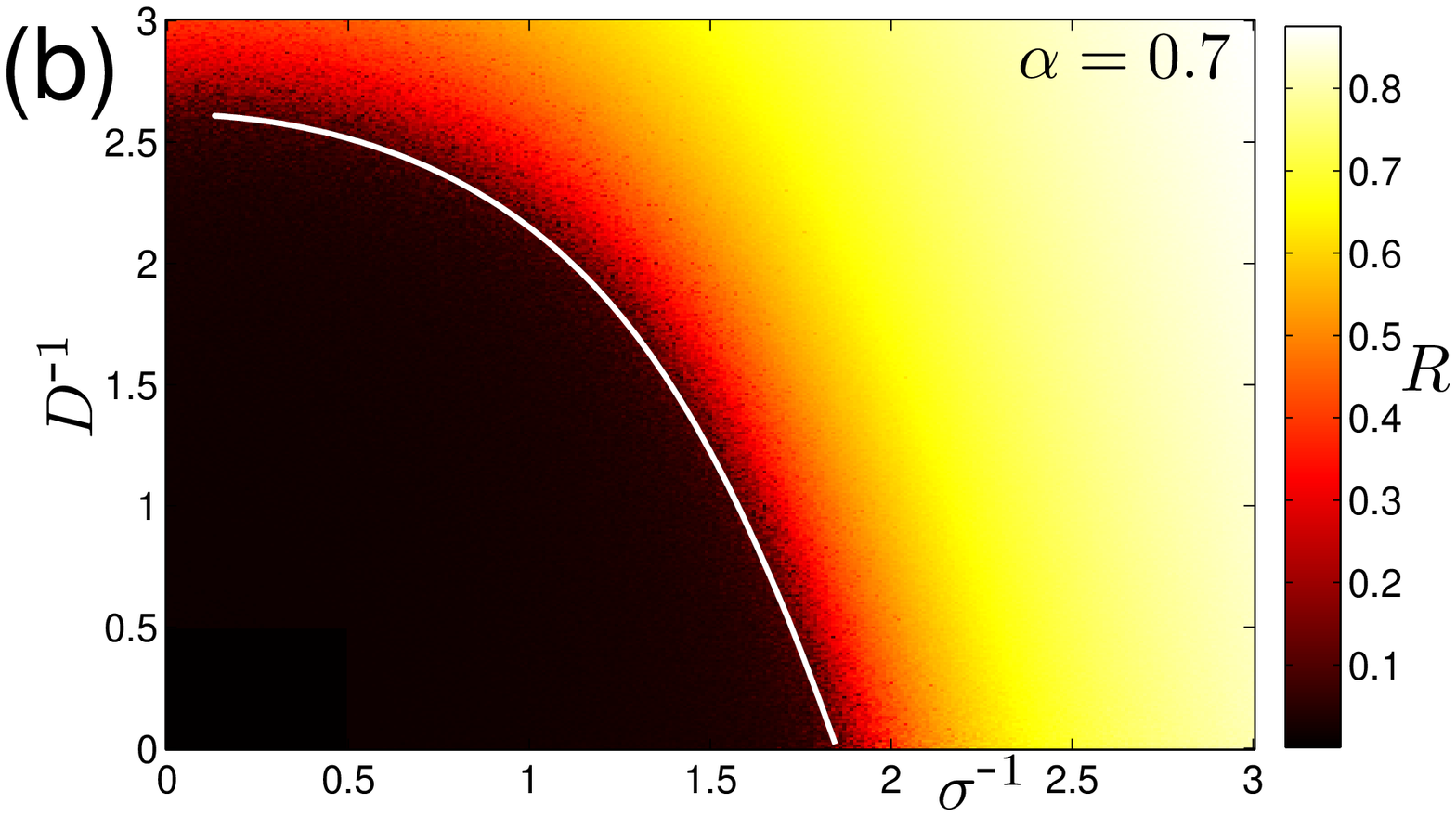}
\caption{\small (color online) Synchronization transition of the Kuramoto model Eq.~(\ref{Eq:KM01}) subject to random forces of Ornstein-Uhlenbeck type with variance $\sigma^2$ and dissipation rate $\gamma$. Shown is the time averaged order parameter $R$ as a function of $\sigma^{-1}$ and $D^{-1}=\gamma/\sigma^2$ for (a) $\alpha=0.0$ and (b) $\alpha=0.7$ in a system of size $N=5000$ averaged over $150$ time units after a transient of $50$ units. The solid white line marks the critical conditions Eq.~(\ref{Eq:xyRelations}) obtained by changing $x=D^{-1}\sigma$ from $0.01$ to $20$. The white noise limit is located on the ordinate axis for $\sigma^{-1}\to 0$ and the quenched noise limit on the abscissa axis for $D^{-1}\to 0$.}
\label{Fig:OUfig}
\end{figure}
With $\textnormal{i}g_1^*=-\exp(\textnormal{i}\alpha)/2$ it follows that 
\beq	\label{Eq:xyRelations}
	\sigma^{-1} = 2|\mu_0(x,y)-\textnormal{i}y| ~,\textnormal{and}~~\cos\alpha =  -2\sigma\textnormal{Re}\mu_0.
\eeq
The critical lines in Fig.~\ref{Fig:OUfig} are parametrized by the time scale ratio $x=0\dots\infty$. For fixed nonzero $\alpha$ the time scale ratio $y(x)$ has to be determined numerically from Eq.~(\ref{Eq:xyRelations}).
For vanishing $\alpha$, $\mu_0$ is real, i.e. $y=0$. The white noise limit $D_{\textnormal{cr}}=0.5 \cos\alpha$ and $\Omega=0.5 \sin\alpha$ can easily be obtained from Eq.~(\ref{Eq:Kette01}) letting $\sigma\to\infty$. The quenched noise limit $x\to 0$ is not trivial. For $\alpha=0$ it must be compared to $\sigma_{\textnormal{cr}}^{-1}=2/\pi\varphi_0(0) = \sqrt{8/\pi}$ \cite{Kuramoto84}. No simple expression exists for $\alpha\ne 0$ \cite{SakKura86}. As a special case of Eq.~(\ref{Eq:Kette01}), for $\alpha=0$ and $x=1$, we obtain $D_{\textnormal{cr}}^{-1} = \sigma_{\textnormal{cr}}^{-1} = 2/(e-1)$ \cite{Euler1782}. 
\\
To test our analytic result for the critical condition Monte-Carlo simulations of the Kuramoto model Eq.~(\ref{Eq:KM01}) with OU random forces have been carried out with finite step size $dt$. The displacements of phase $\vartheta$ and force $\eta$ can be drawn directly from the transition probability $p(\vartheta+d\vartheta,\eta+d\eta,t+dt|\vartheta,\eta,t)$ under the assumption of a slowly changing mean field force $f(\vartheta)$ which is assumed constant during an integration step. The two random variables
\beqarr	\label{Eq:OURanDispl}
	r_1 &=& d\vartheta -f(\vartheta)dt + D\sigma^{-1} (1-e^{-\sigma^2D^{-1}dt})\eta(t),	\nonumber \\ \\
	r_2 &=& \eta(t+dt) - e^{-\sigma^2D^{-1}dt} \eta(t)		\nonumber
\eeqarr
are Gaussian  \cite{Risken89} with correlation matrix
\beqarr	\label{Eq:OUCorr}
	\Sigma_{11} &=& D^2\sigma^{-2} (2\sigma^2D^{-1}dt - 3 + 4e^{-\sigma^2D^{-1}dt} - e^{-2\sigma^2D^{-1}dt}),	\nonumber	\\ \nonumber \\
	\Sigma_{12} &=& \Sigma_{21} ~=~ D\sigma^{-1} (1-e^{-2\sigma^2D^{-1}dt})^2,								\\ \nonumber \\
	\Sigma_{22} &=& 1 - e^{-2\sigma^2D^{-1}dt}									\nonumber
\eeqarr
Both values in Eq.~(\ref{Eq:OURanDispl}) can be sampled at all time scales and in particular also for $\sigma^{-1}\to 0$ as well as $D^{-1}\to 0$. The critical line obtained from Eq.~(\ref{Eq:Kette01}) and the numerical simulations agree very well (Fig.~\ref{Fig:OUfig}).
\\
\section*{Conclusions}
%
%
By means of linear stability analysis we have succeeded to find critical conditions for the transition to synchronization in the Kuramoto model of globally coupled, identical oscillators subject to independent but identically distributed colored noise forces in the cases of Ornstein-Uhlenbeck type and random Telegraph noise. We are hopeful that our results can be applied to obtain qualitative and quantitative predictions for the critical coupling strength in an ensemble of phase coherent chaotic oscillators \cite{PiRoKu96,Sakaguchi00} or networks of identical autonomous oscillators \cite{ToKoMa10}. For such an application it will be necessary to approximate the experimentally accessible fluctuations in single oscillators by a linear model such as the Ornstein-Uhlenbeck process or a finite state Markov model like the random telegraph process.
\\ \\
Ther author thanks H. Kori for valuable feedback.
This work was supported by JST Special Coordination Funds for Promoting Science and Technology.

\end{document}